\documentclass[12pt,a4paper]{article}
\usepackage{amsmath,amssymb,enumerate,bbm,color,alltt,url}
\usepackage{hyperref}
\usepackage{cleveref}
\usepackage[final]{graphicx}
\usepackage{etex}
\reserveinserts{50}
\usepackage{morefloats}
\usepackage{color}
\usepackage{verbatim}
\usepackage{float}
\usepackage{subfig}
\usepackage{setspace}
\usepackage{anonchap}
\usepackage{microtype}
\usepackage[utf8]{inputenc}
\usepackage[T1]{fontenc}
\usepackage{lmodern}
\usepackage[final]{listings}
\usepackage{xcolor}

\title{
A MUSTA-FORCE algorithm for solving partial differential equations of relativistic hydrodynamics}

\author{J.~Porter-Sobieraj$^1$,M.~S{\l}odkowski$^2$, D.~Kiko{\l}a$^2$,J.~Sikorski$^3$ \\ 
P.~Aszklar$^1$}

\date{}

\begin{document}

\maketitle

\hspace{-0.6cm}$^1$ Warsaw University of Technology, Faculty of Mathematics and Information Science, Koszykowa 75, 00-662 Warsaw, Poland \\
$^2$ Warsaw University of Technology, Faculty of Physics, Koszykowa 75, 00-662 Warsaw, Poland\\
$^3$ University of Warsaw, Faculty of Physics, Ho\.za 69, 00-681 Warsaw, Poland,\\

\begin{abstract}
Understanding event-by-event correlations and fluctuations is crucial for 
the comprehension of the dynamics of heavy ion collisions. Relativistic 
hydrodynamics is an elegant tool for modeling these phenomena; however, such 
simulations are time-consuming, and conventional CPU calculations are not 
suitable for event-by-event calculations. This work presents a feasibility 
study of a new hydrodynamic code that employs graphics processing units 
together with a general MUSTA-FORCE algorithm (Multi-Stage Riemann 
Algorithm - First Order Centered scheme) to deliver a high-performance yet 
universal tool for event-by-event hydrodynamic simulations. We also 
investigate the performance of selected slope limiters that reduce 
the amount of numeric oscillations and diffusion in the presence of 
strong discontinuities and shock waves. The numerical results are compared 
to the exact solutions to assess the code's accuracy.
\end{abstract}

{\bf Keywords:} relativistic hydrodynamic, simulation of heavy ion collisions, \hspace*{18pt}quark-gluon plasma, high energy nuclear physics, numerical
algorithms, 
\hspace*{18pt}MUSTA-FORCE, parallel computing, CUDA/GPU

\section{Motivation}

Relativistic hydrodynamics simulations are widely used in the modeling of
nuclear processes in high-energy nuclear physics when examining the
properties of the quark gluon plasma (QGP). Detailed information regarding
the reactions that take place at the microscopic level is not required.
Hydrodynamic model formalism treats QGP as a perfect fluid and
assumes a single equation of state. The bulk and hot nuclear system can
be described using hydrodynamic conservation laws and then solved
numerically~\cite{NumProb:PerfFluid1,NumProb:PerfFluid2}. 
Hydrodynamic models are extremely successful in describing
experimental results for particles with low transverse momentum which
is a behavior of bulk nuclear matter.
On the other hand, jets (narrow spays of hadrons and other particles
produced by the hadronization of a high energy quark or gluon) are
widely used to probe the properties of the QGP. This
is an approach analogous to tomography: an external, penetrating
probe, whose properties (like a production mechanism) are under
experimental and theoretical control, is shot through the medium. We
can then infer the properties of the analyzed system from the modification of
the probe's energy. Jets are such probes - external to the
QGP. Because their production requires a large momentum transfer, they
are produced very early in the collision, in the initial hard
interaction, before the QGP phase. Their production is described well
by perturbative quantum chromodynamics (pQCD) calculations. Thus they are 
excellent tools for QGP research.

The mechanism involved in energy loss due to interactions with nuclear
matter has been a topic of extensive theoretical and experimental studies
over the last two decades. 
However, the energy dissipated by jets can also alter the properties of
bulk nuclear matter (for instance, so-called elliptic flow) in the
intermediate transverse momentum range. There is little
understanding of such effects. For such studies, we need to efficiently 
model the soft particle evolution with a high spatial
resolution to capture the jet-induced modification of the
characteristics of the bulk nuclear matter. Moreover, the Cartesian
coordinate system is preferred to ensure a high spatial resolution
that is constant throughout the evolution of the system. Such
calculations are necessary to fully understand the proprieties of this
unique state of nuclear matter.

The main motivation of our work is therefore to develop a new application to
study the physics of heavy ion collisions, enabling the execution of
hydrodynamic simulations on high-resolution Cartesian grids. 
The goal is to combine two areas of heavy ion physics that are usually 
treated separately: bulk physics, described by hydrodynamic models, 
and the physics of jets. Such an approach will allow us to investigate in detail 
how energy deposited by jets affects the behavior of soft particles. 
In turn, it will help to understand better the properties of nuclear 
matter under extreme conditions.

Moreover, event-by-event correlations and fluctuations play a significant role 
in the understanding of heavy ion collision dynamics. Unfortunately, 
hydrodynamic simulations are time- and computing-power-consuming, and 
event-by-event calculations are challenging when conventional CPU computing 
is used. The achievements of hydrodynamic models in high-energy physics 
motivated us to work on a new high-performance program that facilitates 
event-by-event simulation with high numerical precision using 
graphics processing units.

Equations of ideal relativistic hydrodynamics in one spatial dimension 
can be solved both analytically and numerically as shown
e.g. in~\cite{NumProb:Schneider1993,NumProb:Balsara1994,NumProb:Cheng2010}.
On a three dimensional grid, however, an analytical solution is difficult 
to achieve and the problem
needs to be addressed via numerical algorithms. 
A number of such algorithms already exist
solving relativistic hydrodynamic equations for ideal and viscous fluid models 
either directly~\cite{NumProb:Akamatsu2014, NumProb:Karpenko2014, NumProb:Tachibana2014, NumProb:Bozek2012, NumProb:Akkelin2008} or retrieving the solution from a particle-based kinematic 
model~\cite{NumProb:Sagert2014}. These solutions, however, incur a heavy 
computational cost in 3+1
dimensions. Hence, 
to ensure reasonable performance, the simulation is often done in reduced
dimensionality, or one of the dimensions (usually rapidity) is represented in 
low resolution, by only a few layers. This approach is motivated by certain 
symmetries that areapproximately held in heavy ion collisions. 
Full (3+1)-dimensional, high resolution~simulations are currently 
very time-consuming computations
for traditional, single-threaded software.

This problem can be solved by employing general purpose computing on graphics 
processing units (GPGPU). 
The paper \cite{NumProb:Gerhard2013} has shown that a GPGPU-based implementation
of the SHASTA algorithm \cite{NumProb:Boris1973} can provide
up to two orders of magnitude improvement in performance. In addition, 
our previous work, based on the universal multi-stage (MUSTA-FORCE) algorithm 
\cite{NumProb:Toro2003,NumProb:Toro2006} also proved that hydrodynamic 
computations can be performed 
within a reasonable timeframe. Although the implemented method has a 
high numerical cost and
complexity, it scales perfectly with parallel computations. Our implementation 
turned out to be over 200 times faster than a sequential implementation on a CPU
~\cite{NumProb:FedCSIS2013}.
It produces good statistics and high spacial and temporal resolutions at the 
same time.  

The speed-up gained by using a GPU made it possible to develop a valuable 
new tool for high energy nuclear science. Our current GPU implementation 
allows a device to perform a massive number of simulations, and furthermore, 
the MUSTA-FORCE algorithm is a very universal tool. Its strength lies in the 
fact that it uses simple 
central schemes and does not require any knowledge of the physical process's 
details. 
However, in the case of large gradients it is necessary to apply a slope 
limiter to reduce numerical oscillations. Therefore, we focused here on 
how to improve the accuracy of the MUSTA-FORCE algorithm. 
The hydrodynamic simulation results presented in this paper show that 
the MUSTA-FORCE method 
is very sensitive to the choice of slope and slope limiter, and 
the way that it is applied. 
We used such a procedure for each dimension separately, but in general 
it could be applied in other ways (e.g. a common slope limiting value could be 
chosen for all the variables). There is no
general procedure in a multi-dimensional and non-scalar case, and always 
some experimentation is necessary for both a particular system of equations, 
and perhaps even a particular problem. 
Under the conditions of our simulation the schemes must be especially 
sensitive to problems containing both strong discontinuities and smooth 
solution features.

The paper is organized as follows. The MUSTA-FORCE algorithm, together 
with the slope limiters, is described in detail in Section~2. 
Section~3 presents the results of using such slope limiters for standard 
benchmarks in nuclear physics and compares them with known analytical solutions.
We discuss their 
ability to achieve high order accuracy in smooth regions while maintaining 
stable, non-oscillatory and sharp discontinuity transitions. The last section 
concludes our paper.

\section{Hydrodynamics Simulations}
\subsection{Mathematical Description}

Relativistic hydrodynamics simulations are based on hyperbolic partial differential equations in the form:
\begin{equation}
	\frac{\partial U}{\partial t} + \frac{\partial F ( U)}{\partial x} +
	\frac{\partial G ( U)}{\partial y} + \frac{\partial H ( U)}{\partial z} = 0
	\label{eqn:hydro1}
\end{equation}
and an equation of state: 
\begin{equation}
	p = p ( e, n)
	\label{eqn:hydro2}
\end{equation}
$U = ( E, M_x, M_y, M_z, R)$ is a vector of conserved quantities in the 
\textit{laboratory rest frame};  $E$ is the energy density, $M_x$, $M_y$ and $M_z$ are the momentum densities 
in the $x$, $y$ and $z$ Cartesian coordinates, respectively, and $R$ is a conserved charge density. 
Vectors of fluxes $F$, $G$, $H$ in the $x$, $y$, and $z$ directions are defined as:
\begin{equation}
  \begin{array}{c}
	  F ( U) = \left[\begin{array}{c}
	    ( E + p) v_x\\
	    M_x v_x + p\\
	    M_y v_x\\
	    M_z v_x\\
	    Rv_x
	  \end{array}\right] \\ \\
	    G ( U) = \left[\begin{array}{c}
	    ( E + p) v_y\\
	    M_x v_y\\
	    M_y v_y + p\\
	    M_z v_y\\
	    Rv_y
	  \end{array}\right] \\ \\
	  
	  H ( U) = \left[\begin{array}{c}
	    ( E + p) v_z\\
	    M_x v_z\\
	    M_y v_z\\
	    M_z v_z + p\\
	    Rv_z
	  \end{array}\right] \label{eqn:fluxes}
  \end{array}
\end{equation}
where $v$ is the velocity, and $p$ is pressure, defined by the energy $e$ and charge density $n$ in the \textit{fluid rest frame}, where velocity $v$ vanishes ($v = ( 0, 0,0)$). Additionally, the following relations occur:
\begin{eqnarray}
  E & = & (\varepsilon + p) \gamma^2 - p \nonumber\\
  M_i & = & (\varepsilon + p) \gamma^2 v_i, \hspace{2em} i = x, y, z \label{eqn:EMR} \\
  R & = & n \gamma \nonumber
\end{eqnarray}
where $\varepsilon$ is the energy density and $\gamma = \tfrac{1}{\sqrt{1-v^2}}$ is the Lorentz factor. Eq.~\ref{eqn:EMR} defines the transformation from \textit{rest frame} variables to \textit{conserved} variables used in integration.

\subsection{Initial Conditions}

To start hydrodynamic evolution, an initial state is required as input.

The most basic is parametrizations based e.g. on Glauber-like models in the transverse plane
(see~\cite{NumProb:IniCond1} for a review), and Bjorken's solution in the longitudinal direction.

Other approaches involve models based on color glass condensate (CGC),
which describe a Lorentz contracted and slowed down, fast moving particle; 
pQCD+saturation model~\cite{NumProb:IniCond2}, or the string rope model~\cite{NumProb:IniCond3}.

These models describe a smooth, averaged initial state. However, since the hydrodynamic equations
are nonlinear, a solution with an averaged initial state is not equivalent to the average of solutions
with fluctuating initial conditions. Because of this, event-by-event calculations became a major
point of interest.

Fluctuating initial conditions can be obtained using e.g. Monte--Carlo Glauber \cite{NumProb:IniCond4,NumProb:IniCond5,NumProb:IniCond9} or CGC, SPheRIO, NeXus~\cite{NumProb:IniCond6}, NeXSPheRIO~\cite{DerradideSouza:2011rp}, and models like EPOS~\cite{NumProb:IniCond7} or UrQMD~\cite{NumProb:IniCond8}.

\subsection{Time Integration}

For time propagation the standard Runge--Kutta methods are employed \cite{NumProb:RK}. 
For numerical stability only total variation diminishing methods are used.

In general, a Runge--Kutta method for Eq.~\ref{eqn:hydro1} can be written in
the form:
\begin{eqnarray}
  U_{(0)}^n & = & U^n \nonumber\\
  U_{(i)}^n & = & \sum^{i - 1}_{k = 0} (\alpha_{ik} U_{(k)}^n + \Delta t
  \beta_{ik} L (U_{(k)}^n)), \nonumber\\
  & &i = 1, \ldots, m \\
  U^{n + 1} & = & U_{(m)}^n \nonumber
\end{eqnarray}
where the upper index without parentheses denotes the time step, the lower index
denotes integration step, $L$ is a numerical recipe to calculate the negative
flux gradient in Eq.~\ref{eqn:hydro1} and $\alpha, \beta$ are constant coefficients
given for a particular method.

For second order accuracy the following method is used:
\begin{eqnarray}
  U_{(1)}^n & = & U^n + \Delta tL (U^n) \nonumber\\
  U^{n + 1} & = & \frac{1}{2}  (U^n + U_{(1)}^n + \Delta tL (U_{(1)}^n)) 
\end{eqnarray}
and for third order accuracy:
\begin{eqnarray}
  U_{(1)}^n & = & U^n + \Delta tL (U^n) \nonumber\\
  U_{(2)}^n & = & \frac{3}{4} U^n + \frac{1}{4} U_{(1)}^n + \frac{1}{4} \Delta tL
  (U_{(1)}^n) \\
  U^{n + 1} & = & \frac{1}{3} U^n + \frac{2}{3} U_{(2)}^n + \frac{2}{3} \Delta
  tL (U_{(2)}^n) \nonumber
\end{eqnarray}
It is apparent that apart from the additional computational cost due to more
evaluations of $L$, these methods introduce the need for an additional storage
register for each of the conserved variables. As this can be an issue for
large resolution simulations, a low storage version of the third order method
can be used.

\subsection{Hybrid MUSTA-FORCE Algorithm}

To obtain a general and accurate solution for Eq.~\ref{eqn:hydro1} and Eq.~\ref{eqn:hydro2}, 
we use a hybrid MUlti--STAge (MUSTA) approach~\cite{NumProb:Toro2006,NumProb:Toro1999}. 
This utilizes a centered flux in a predictor--corrector loop, solving the Riemann problem 
numerically, i.e. without using a priori information about waves.

In order to calculate flux $F_{i + \frac{1}{2}}$ the algorithm, in a one dimensional case, is as follows:
\begin{enumerate}
  \item Introduce auxiliary variables $U^{(l)}_L$ and $U^{(l)}_R$ and 
  their fluxes $F^{(l)}_L$ and $F^{(l)}_R$.
  
  \item Set $U^0_L = U_i$, $U^0_R = U_{i + 1}$.
  
  \item Calculate $F^{(l)}_{i + \frac{1}{2}}$ using a centered flux,
  $U^{(l)}_L$, $U^{(l)}_R$, $F^{(l)}_L$ and $F^{(l)}_R$. If $l$ reached a
   maximum number of iterations, stop.
  
  \item Solve Riemann problem locally:
  \begin{equation}
	   \begin{array}{c}
	    U^{( l + 1)}_L = U^{( l)}_L - \frac{\Delta t}{\Delta x} \left( F^{( l)}_{i
	    + \frac{1}{2}} - F^{( l)}_L \right), \\
	    U^{( l + 1)}_R = U^{( l)}_R -
	    \frac{\Delta t}{\Delta x} \left( F^{( l)}_R - F^{( l)}_{i + \frac{1}{2}}  \right).
    \end{array}
  \end{equation}
  \item Go back to step 3.
\end{enumerate}
As a centered flux in step 3, we use the First ORder CEntered (FORCE) scheme:
\begin{equation}
  F^{\rm{force}}_{i + \frac{1}{2}} = \frac{1}{2} \left( F^{\rm{lw}}_{i +
  \frac{1}{2}} + F^{\rm{lf}}_{i + \frac{1}{2}} \right)
\end{equation}
where $F^{\rm{lw}}_{i + \frac{1}{2}}$ is the Lax-Wendroff type flux:
\begin{equation}
  F^{\rm{lw}}_{i + \frac{1}{2}} = F \left( \frac{1}{2}  ( U_L + U_R) -
  \frac{\alpha \Delta t}{2\Delta x}  ( U_R - U_L) \right)
\end{equation}
and $F^{\rm{lf}}_{i + \frac{1}{2}}$ is the Lax-Friedrichs type flux:
\begin{equation}
  F^{\rm{lf}}_{i + \frac{1}{2}} = \frac{1}{2}  ( F_L + F_R) -
  \frac{\Delta x}{2\alpha \Delta t}  ( U_R - U_L)
\end{equation}
In a three-dimensional case $\alpha = 3$, but other values may also be
considered.

To achieve second order accuracy in space and time, we extend our algorithm 
with the MUSCL-Hancock scheme. The basic idea of this scheme is to use more cells 
to interpolate inter-cell values and evolve them half a time step:
\begin{enumerate}
  \item Replace cell average values $U_i^n$ by a piecewise linear function
  inside $i$-th cell:
  \begin{equation}
    U_i ( x) = U^n_i + \frac{( x - x_i)}{\Delta x} \Delta_i \label{eqn:muscl1}
  \end{equation}
  where $\Delta_i$ is a slope vector and will be defined later.
  
  In the local coordinates the points $x = 0$ and $x = \Delta x$ correspond to
  boundaries of the cell $x_{i - \frac{1}{2}}$ and $x_{i + \frac{1}{2}}$. The
  values at these points are $U^L_i = U^n_i -\Delta_i/2$ and $U^R_i
  = U^n_i + \Delta_i /2 $.
  
  \item Propagate $U^L_i$ and $U^R_i$ by a time $\frac{1}{2} \Delta t$:
\end{enumerate}
  \begin{eqnarray}
    \tilde{U}^L_i & = & U^L_i + \frac{1}{2}\frac{\Delta t}{\Delta x}  ( F (
    U^L_i) - F ( U^R_i)) \nonumber \\
    & & + \frac{1}{2}\frac{\Delta t}{\Delta y}  ( G ( U^L_i) - G ( U^R_i)) \nonumber \\
    & & + \frac{1}{2}\frac{\Delta t}{\Delta z}  ( H ( U^L_i) - H ( U^R_i)) \nonumber \\
    \tilde{U}^R_i & = & U^R_i + \frac{1}{2}\frac{\Delta t}{\Delta x}  ( F (
    U^L_i) - F ( U^R_i)) \\
    & & + \frac{1}{2}\frac{\Delta t}{\Delta y}  ( G ( U^L_i) - G ( U^R_i)) \nonumber \\
    & & + \frac{1}{2}\frac{\Delta t}{\Delta z}  ( H (  U^L_i) - H ( U^R_i)) \nonumber 
  \end{eqnarray}
  \begin{enumerate}
  	\setcounter{enumi}{2}
  \item Use $\tilde{U}^L_i$ and $\tilde{U}^R_i$ as $U^0_L$ and $U^0_R$ in
  MUSTA.
\end{enumerate}

A simple choice for the \textit{slope} $\Delta_i$ in Eq.~\ref{eqn:muscl1} is:
\begin{equation}
  \Delta_i = \frac{1}{2}  ( U^n_{i + 1} - U^n_{i - 1}) \label{eqn:slope}
\end{equation}
which indeed results in a second-order accurate algorithm. However, as predicted 
by Godunov's theorem, it has the unpleasant effect of producing spurious 
oscillations in the case of strong gradients.
Complex study of MUSTA schemes can be found in ~\cite{NumProb:ToroTitarev2006}.

\subsection{Slope Limiters in the MUSTA-FORCE}

To avoid such oscillations and to solve problems that appear in the presence of shocks, discontinuities 
or sharp changes, flux limiting and slope limiting methods have
been proposed \cite{NumProb:Slope1,NumProb:Slope2}.
We employed a slope limiting method; instead of $\Delta_i$ as in Eq.~\ref{eqn:slope} we use 
\begin{equation}
\tilde{\Delta}_i = \xi (r_i) (U_{i} - U_{i - 1})
\end{equation}
in Eq.~\ref{eqn:muscl1}, where $\xi$ is called the slope limiter and
\begin{equation}
r_i = \frac{U_{i+1} - U_{i}}{U_{i} - U_{i-1}}.
\end{equation}

Then one can calculate $U_i^L$ and $U_i^R$ using the following relations

  \begin{equation}
    \begin{array}{c}
     U_i^L = U_i - \frac{1}{2} \xi(1/r_i) (U_{i+1} - U_i), \\
     U_i^R = U_i + \frac{1}{2} \xi(r_i) (U_{i} - U_{i-1}).
    \end{array}
  \end{equation}

There are a number of possible choices for $\xi$, each with its own characteristics and features. The four different limiters investigated in this paper are Minbee (MB) \cite{NumProb:Minbee}, Superbee (SB) \cite{NumProb:Superbee}, van Albada (VA) \cite{NumProb:vanAlbada} and van Leer (VL) \cite{NumProb:vanLeer}.
They are expressed as:
  \begin{equation}
  	\xi_{\rm{MB}} (r) = \max (0, \min (1, r))
  \end{equation}
  \begin{equation}
  	\xi_{\rm{SB}} (r) = \max (0, \min (2 r, 1), \min (r, 2))
  \end{equation}
  \begin{equation}
  	\xi_{\rm{VA}} (r) = \frac{r^2+r}{r^2+1} 
  \end{equation}
  \begin{equation}
	\xi_{\rm{VL}} (r) = \frac{r+|r|}{1+|r|} 
  \end{equation}

All the four limiters are symmetric, i.e. they meet the symmetry property:
  \begin{equation}
	\xi (\frac{1}{r}) = \frac{\xi (r)}{r} 
  \end{equation}
that provides that forward and backward gradients are treated in the same manner.
Otherwise, the results on the left would differ from those on the right despite
initial symmetry in the system.

The limiters are designed to reduce the scheme to first order accuracy near
shocks, and keep higher order in smooth areas. Introducing non--linearity in
this way reduces spurious oscillations and retains good accuracy of the
solution.

The two most extreme slope limiters are Minbee and Superbee. The first one is
the most dissipative and the second - the least. Between them lies the
admissible region for second order total variation diminishing (TVD) limiters.

Here it should also be stressed, that results are very sensitive to the choice of
slope in Eq.~\ref{eqn:slope} and the slope limiter---both the formula for $\xi$, 
and the way that it is applied.

\subsection{Calculating the hydrodynamic flux}

To complete the chapter about numerical methods for relativistic hydrodynamics,
one more detail must still be dealt with. It is easy to change from rest frame variables and velocities
to the conserved variables used in the integration. The inverse transformation, however, is not trivial.

It is needed, however, each time we compute the flux, since the equation of state (\ref{eqn:hydro2}) 
is defined as a function of $e$ and $n$, and not $E$ and $R$. As in \cite{NumProb:Rischke1999}, 
we can invert equations~(\ref{eqn:EMR}) and get
\begin{eqnarray}
e &=& E - M v \nonumber \\
n &=& R \sqrt{1-v^2} \label{eqn:vel}\\
v &=& \frac{M}{E + p(e,n)} \nonumber\\
  &=& \frac{M}{E+p(E-M v, R \sqrt{1-v^2})} \nonumber
\end{eqnarray}
The set of equations (\ref{eqn:vel}) can be solved numerically, starting from the last one 
as the fixed point equation for $v$.

\section{Discussion and Results}
\begin{figure*}[hb]
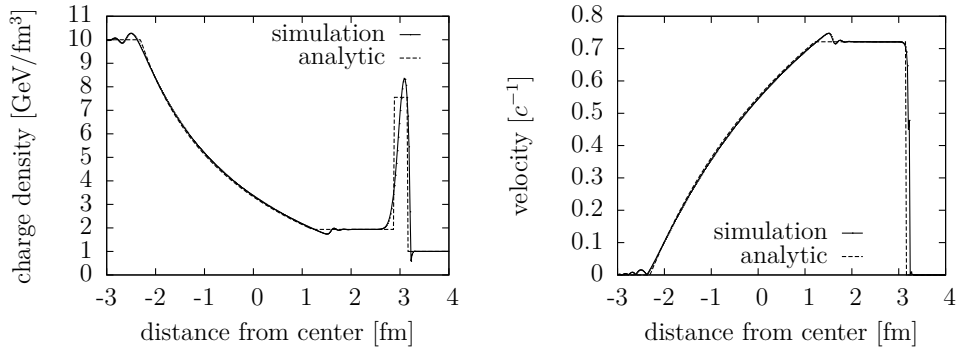

\centering
\scalebox{.8}{\input{./shock_n_mf-none}}
\scalebox{.8}{\input{./shock_v_mf-none}}
\caption{\label{fig:fig1}Sod shock tube, charge density in the local rest frame of the fluid (left) and velocity (right), MUSTA-FORCE with no limiter.}
\end{figure*}

\begin{figure*}[hb]
\centering
\scalebox{.8}{\input{./shock_n_mf-minbee}}
\scalebox{.8}{\input{./shock_v_mf-minbee}}
\caption{\label{fig:fig2}Sod shock tube, charge density in the local rest frame of the fluid (left) and velocity (right), MUSTA-FORCE with with Minbee limiter.}
\end{figure*}

\begin{figure*}[hb]
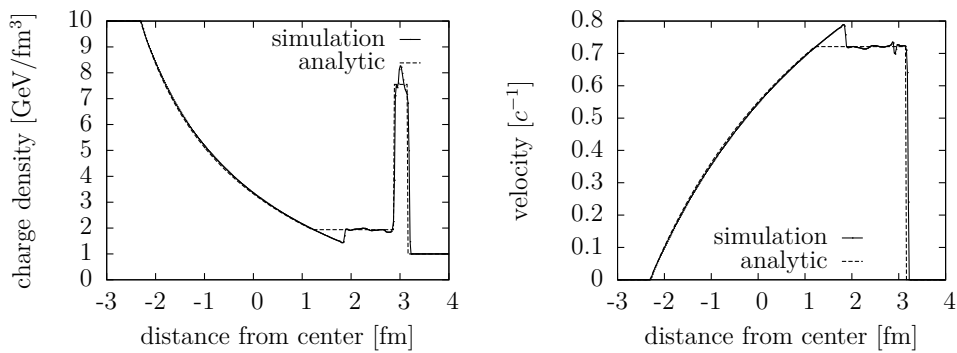

\centering
\scalebox{.8}{\input{./shock_n_mf-superbee}}
\scalebox{.8}{\input{./shock_v_mf-superbee}}
\caption{\label{fig:fig3}Sod shock tube, charge density in the local rest frame of the fluid (left) and velocity (right), MUSTA-FORCE with Superbee limiter.}
\end{figure*}

\subsection{Implementation Notes}

In the case of hydrodynamics simulations, fully (3+1)-dimensional simulations in space and in time 
are much-desired, to describe the system's evolution without any assumptions regarding 
its symmetries and without decreasing the dimension of the problem.
Moreover, event-by-event calculations become a major point of interest, 
with fluctuating initial conditions and with a large amount of statistics.
Such simulations are extremely expensive in terms of computing power 
and require an efficient and fast computer code.

Therefore, we implemented the hydrodynamics simulation algorithm on a GPU using an NVIDIA CUDA framework~\cite{NumProb:CUDA}. 
Due to the large numerical grids in the simulations and the complexity of the computations, 
we used surface memory to store the simulation data. The state of the system is saved 
as 5 single precision floating-point numbers per lattice cell -- the energy density, 
conserved charge density, and 3 momenta density; all in the \textit{laboratory reference frame}. 
The maximum grid that fits then within the surface memory limitations is $240^3$. 

\subsection{Numerical Experiments}

To verify the simulation reliability, the MUSTA-FORCE algorithm itself was tested (along with 
the four slope limiters -- Minbee, Superbee, van Albada and van Leer) 
against three analytical solutions to relativistic hydrodynamics -- 
the Sod shock tube \cite{NumProb:Thompson1986,NumProb:Marti2003}, the Hubble-like expansion \cite{NumProb:Chojnacki2005} and ellipsoidal flow \cite{NumProb:Sinyukov2005,NumProb:Sinyukov2006}. 
For each of them, plots of chosen variables are presented together with the theoretical curves.

The number of stages and the order of Runge--Kutta method were fitted experimentally. 
In all the presented cases four MUSTA stages guaranteed stability and good numerical accuracy in 
acceptable calculation time on GPU. The third order accurate Runge--Kutta method was used 
for time integration.

Many numerical experiments for different parameters were carried out to
verify correctness, convergence and robustness of the MUSTA-FORCE method. 
We present here the results for one representative set of parameters 
per analytical solution. The initial conditions and parameters 
like time step $\Delta t$ and grid spacing $\Delta x$, $\Delta y$ and $\Delta z$
were given for each experiment separately. In all the cases the Courant number 
was less than 1. More detailed analysis of CFL numbers and stability limits in 
the MUSTA approach can be found in~\cite{NumProb:Blakely}.

\subsubsection{Sod Shock Tube}

The first test is a solution to the Riemann problem. This is a one dimensional solution, whose
initial state comprises two regions of stationary fluid with a charge and pressure discontinuity in the middle.

When the discontinuity is big enough, a relativistic shock wave appears in the solution.
The initial conditions (given in Table~\ref{tbl:shockpars} together with other parameters) 
were chosen to produce such a shock wave.  

\begin{figure*}[ht]
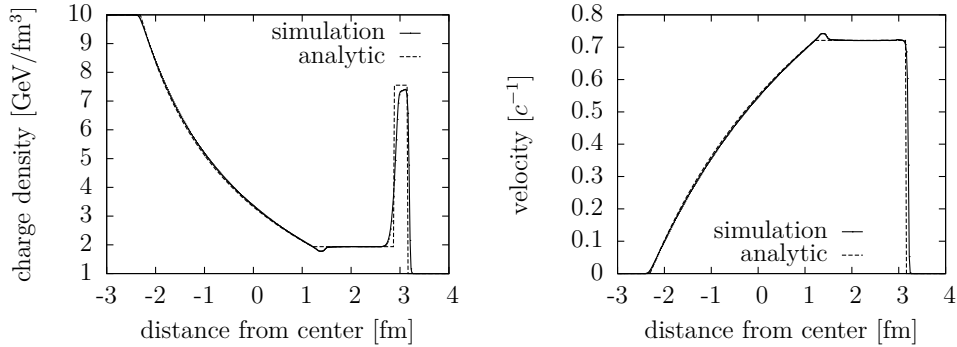

\centering
\scalebox{.8}{\input{./shock_n_mf-albada}}
\scalebox{.8}{\input{./shock_v_mf-albada}}
\caption{\label{fig:fig4}Sod shock tube, charge density in the local rest frame of the fluid (left) and velocity (right), MUSTA-FORCE with van Albada limiter.}
\end{figure*}

\begin{figure*}[ht]
\centering
\scalebox{.8}{\input{./shock_n_mf-leer}}
\scalebox{.8}{\input{./shock_v_mf-leer}}
\caption{\label{fig:fig5}Sod shock tube, charge density in the local rest frame of the fluid (left) and velocity (right), MUSTA-FORCE with van Leer limiter.}
\end{figure*}

\begin{table}[h]
	\centerline{
	\begin{tabular}{|c|c|}
		\hline
		parameter & value \\
		\hline
		grid size & $500$ \\
		grid spacing & $0.02$ \\
		time step & $0.005$ \\
		$p_L$ & $13 \frac{1}{3}$ \\
		$p_R$ & $0$ \\
		$n_L$ & $10$ \\
		$n_R$ & $1$ \\
		\hline
	\end{tabular}}
  	\caption{Parameters of the Sod shock tube simulations \label{tbl:shockpars}}
\end{table}


The solution is divided into waves: the shock wave, the contact discontinuity, and the rarefaction wave.

The results are presented in Figs.~\ref{fig:fig1}--\ref{fig:fig5}.

For the \textit{no limiter} case (Fig.~\ref{fig:fig1}) with the MUSTA-FORCE algorithm, oscillations at the contact points of waves become visible. The shock is also smeared out from the left side. 
For Minbee (Fig.~\ref{fig:fig2}) the overshoot is almost entirely gone, at the cost of some visible diffusion, 
especially in the shock wave region.
Superbee in Fig.~\ref{fig:fig3} is the most compressive limiter. The shocks are sharpened, but additional oscillations are introduced. Also the overshoot clearly visible in the velocity plot is enhanced.
The van Albada limiter in Fig.~\ref{fig:fig4} trades some sharpness for better rendition of the velocity profile; f
and for van Leer (Fig.~\ref{fig:fig5}) the shocks are sharp and the oscillations are gone, but the overshoot is still significant. 

To sum up, van Albada and the Minbee limiter seem to be closest to the analytic solution, 
and those two will be presented in rest of the tests.

\subsubsection{Hubble-like Expansion}

This is a three-dimensional, spherically symmetrical solution of matter that expands uniformly.
The velocity is proportional to the distance from the center $v = \frac{\vec{r}}{t}$. The energy
density is given by:
\begin{equation}
	e = e_0 \left( \frac{\tau_0}{\sqrt{t^2-r^2}} \right)^{3 (1+c_s^2)}
\end{equation}
In our case the solution is well defined for $r < t$. For the test we set $r < t - 0.5$ fm
and put a vacuum ($e = v = 0$) outside this region.
This means that the solution is exact only in the central area---on the periphery the matter
will expand into the vacuum, so a rarefaction wave is expected.
The solution uses an ultra-relativistic equation of state $p = c_s^2 e$.

Initial parameters are given in Table~\ref{tbl:hubblepars}.

\begin{table}[h]
	\centerline{
	\begin{tabular}{|c|c|}
		\hline
		parameter & value \\
		\hline
		grid size & $120^3$ \\
		grid spacing & $0.1$ \\
		time step & $0.03$ \\
		$e_0$ & $1$ \\
		$c_s^2$ & $\frac{1}{3}$ \\
		$\tau_0$ & $4$ \\
		$t_0$ & $2$ \\
		\hline
	\end{tabular}}
	\caption{Parameters of the Hubble--like expansion simulations \label{tbl:hubblepars}}
\end{table}


The results are presented in Fig.~\ref{fig:fig6} and Fig.~\ref{fig:fig7}. 
$y = z = 0$ sections are shown through the three dimensional solution.

\begin{figure*}[ht]
\centering
\scalebox{.8}{\input{./hubble_e_mf-minbee}}
\scalebox{.8}{\input{./hubble_v_mf-minbee}}
\caption{\label{fig:fig6}Hubble-like expansion, energy density in the local rest frame of the fluid (left) and velocity (right), MUSTA-FORCE with Minbee limiter.}
\end{figure*}

\begin{figure*}[ht]
\centering
\scalebox{.8}{\input{./hubble_e_mf-albada}}
\scalebox{.8}{\input{./hubble_v_mf-albada}}
\caption{\label{fig:fig7}Hubble-like expansion, energy density in the local rest frame of the fluid (left) and velocity (right), MUSTA-FORCE with van Albada limiter.}
\end{figure*}

The results are similar to those in the previous test. 
The schemes were accurate in the middle region, and
the Minbee limiter has shown less diffusion than van Albada at the sides.

\subsubsection{Ellipsoidal Flow}

\begin{table}[h]
	\centerline{
	\begin{tabular}{|c|c|}
		\hline
		parameter & value \\
		\hline
		grid size & $120^3$ \\
		grid spacing & $0.1$ \\
		time step & $0.02$ \\
		$t_0$ & $2$ \\
		$C_e$ & $2$ \\
		$C_n$ & $0.75$ \\
		$b_e$ & $1$ \\
		$b_n$ & $1$ \\
		$T_1$ & $0.4$ \\
		$T_2$ & $0.6$ \\
		$T_3$ & $0.8$ \\
		\hline
	\end{tabular}}
	\caption{Parameters of the ellipsoidal flow simulations \label{tbl:ellippars}}
\end{table}

\begin{figure*}[ht]
\centering
\scalebox{.8}{\input{./ellip_e_mf-minbee}}
\scalebox{.8}{\input{./ellip_v_mf-minbee}}
\caption{\label{fig:fig8}Ellipsoidal flow, energy density in the local rest frame of the fluid (left) and velocity (right), MUSTA-FORCE with Minbee limiter.}
\end{figure*}

\begin{figure*}[ht]
\centering
\scalebox{.8}{\input{./ellip_e_mf-albada}}
\scalebox{.8}{\input{./ellip_v_mf-albada}}
\caption{\label{fig:fig9}Ellipsoidal flow, energy density in the local rest frame of the fluid (left) and velocity (right), MUSTA-FORCE with van Albada limiter.}
\end{figure*}

\begin{figure*}[ht]
\centering
\input{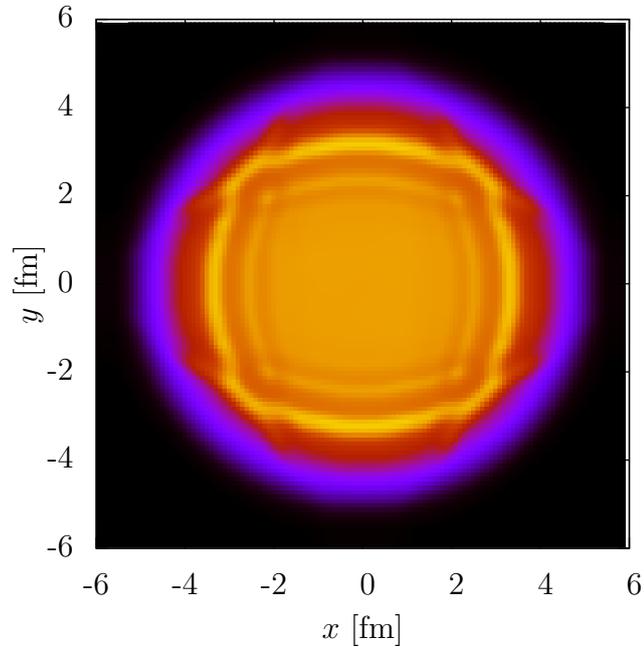}
\caption{\label{fig:fig10}Ellipsoidal flow in the $xy$ plane, energy density in the local rest frame of the fluid, MUSTA-FORCE with van Albada limiter.}
\end{figure*}

The last test uses the ellipsoidal solution, which is a
generalized Hubble--like solution (with velocity proportional to $\vec{r}$), a
gaussian profile and vanishing pressure $p=0$. The variables are given by the following equations:

\begin{eqnarray}
	e &=& \frac{C_e}{\prod\limits_i (t+T_i)}\exp\left(-b_e^2\frac{t^2}{\tau^2} \right)\\
	n &=& \frac{C_n}{\prod\limits_i (t+T_i)}\exp\left(-b_n^2\frac{t^2}{\tau^2} \right)\\
	\vec{v} &=& \left( \frac{a_1(t) x}{t},  \frac{a_3(t) y}{t}, \frac{a_2(t) z}{t}\right)
\end{eqnarray}
where $\tau = \sqrt{t^2 - \sum\limits_i a_i^2 x_i^2}$, $a_i \equiv a_i(t) = t/(t+T_i)$ and
$C_e$, $C_n$, $b_e$, $b_n$, $T_i$ are constants; $i=1,2,3$.

Initial parameters are given in Table~\ref{tbl:ellippars}.


The results are presented in Fig.~\ref{fig:fig8} and Fig.~\ref{fig:fig9}. 
$y = z = 0$ sections are shown through the three dimensional solution.
This solution, despite being pressureless, most 
resembles a physically relevant situation.
The Minbee limiter has shown again less oscillations than others at the sides.

\section{Conclusions}
As a result of this paper, we have implemented and tested a MUSTA-FORCE 
algorithm dedicated to solving  conservative field equations. The universal 
MUSTA-FORCE algorithm proved to be quite efficient and robust. 
Despite promising initial results, it turned out that the MUSTA-FORCE produced 
numerical oscillations. 
Fig.~\ref{fig:fig10} illustrates two dimensional energy density cross section in the $xy$ plane
of the MUSTA-FORCE with Albada limiter. It confirms the clearly visible anisotropy of the numerical oscillations.
To address this issue, we performed a detailed study of universal slope limiters
to reduce oscillations or asymmetric propagation. In all test 
cases the Minbee limiter has shown the best numerical properties.

This implementation of a relativistic hydrodynamic code is designed to run 
efficiently on contemporary graphics processing units,
which have many times more computing power compared to ordinary CPU processors.
Benchmarks and comparison to an equivalent implementation of some of these 
algorithms in C show speedups of over 2 orders of magnitude. 
Thanks to such performance levels, event-by-event analyses can be conducted 
efficiently with a relatively low cost, just a few modern workstations. 
Our work also provides a framework for implementing other algorithms for 
high-statistics event-by-event hydrodynamic simulations.

All the calculations were made as fully (3+1)-dimensional simulations in 
space and in time, without any assumptions regarding its symmetries and 
without decreasing the dimension of the problem. Such (3+1)-dimensional 
numerical simulation on high-resolution Cartesian grids is a novel result. 
It allows us to run hydrodynamic simulations, 
provide information about properties of the relativistic bulk nuclear matter (QGP) by studying  
jet-medium interaction and jet-induced flow. Furthermore event-by-event simulations allow us 
to investigate initial condition fluctuation of the heavy ion collision in the pre-equlibrium  
stage and it might have an impact on the dynamic in the nuclear medium.

\end{document}